\documentclass[conference,10pt]{IEEEtran}
\IEEEoverridecommandlockouts

\usepackage{epsfig,rotating,setspace,latexsym,amsmath,epsf,amssymb,amsfonts,bm,theorem,cite,algorithm,graphicx,epsf,authblk,epstopdf,color,algpseudocode,bbm}

\newtheorem{definition}{Definition}
\newtheorem{theorem}{Theorem}
\newtheorem{lemma}{Lemma}

\allowdisplaybreaks

\begin{document}

\title{Energy Harvesting Networks with General Utility Functions: Near Optimal Online Policies\thanks{This work was supported by NSF Grants CNS 13-14733, CCF 14-22111, CCF 14-22129, and CNS 15-26608.}}

\author{Ahmed Arafa \quad Abdulrahman Baknina \quad Sennur Ulukus\\
\normalsize Department of Electrical and Computer Engineering\\
\normalsize University of Maryland, College Park, MD 20742\\
\normalsize \emph{arafa@umd.edu} \quad \emph{abaknina@umd.edu} \quad \emph{ulukus@umd.edu}}

\maketitle

\begin{abstract}
We consider online scheduling policies for single-user energy harvesting communication systems, where the goal is to characterize online policies that maximize the long term average utility, for some general concave and monotonically increasing utility function. In our setting, the transmitter relies on energy harvested from nature to send its messages to the receiver, and is equipped with a finite-sized battery to store its energy. Energy packets are independent and identically distributed (i.i.d.) over time slots, and are revealed causally to the transmitter. Only the average arrival rate is known a priori. We first characterize the optimal solution for the case of Bernoulli arrivals. Then, for general i.i.d. arrivals, we first show that {\it fixed fraction policies} \cite{ozgur_online_su} are within a constant {\it multiplicative} gap from the optimal solution for all energy arrivals and battery sizes. We then derive a set of sufficient conditions on the utility function to guarantee that fixed fraction policies are within a constant {\it additive} gap as well from the optimal solution.
\end{abstract}

\section{Introduction}

A single-user communication channel is considered, where the transmitter relies on energy harvested from nature to send its messages to the receiver. The transmitter has a battery of finite size to save its incoming energy, and achieves a reward for every transmitted message that is in the form of some general  concave increasing utility function of the transmission power. The goal is to characterize {\it online} power control policies that maximize the long term average utility subject to energy causality constraints.

{\it Offline} power scheduling in energy harvesting communication systems has been extensively studied in the recent literature. Earlier works \cite{jingP2P, kayaEmax, omurFade, ruiZhangEH} consider the single-user setting. References \cite{jingBC, elifBC, omurBC, jingMAC, aggarwalPmax, onurCoopEH, kaya-interference} extend this to broadcast, multiple access, and interference settings; and \cite{ruiZhangRelay, gunduz2hop, letaiefRelay, berkDiamond-jour} consider two-hop and relay channels. Energy cooperation and energy sharing concepts are studied in \cite{berkCoop, kayaProcrastinate}. References \cite{kayaRxEH, yatesRxEH1, payaroRxEH, rahulOnOff, arafaJSACdec, arafa_baknina_twc_dec_proc} study energy harvesting receivers, where energy harvested at the receiver is spent mainly for sampling and decoding. Other works \cite{arafa_baknina_twc_dec_proc, ruiZhangNonIdeal, orhan-broadband, omurHybrid, payaro-cost, baknina_temp} study the impact of processing costs, i.e., the power spent for circuitry, on energy harvesting communications.

Recently, \cite{ozgur_online_su} has introduced an online power control policy for a single-user energy harvesting channel that maximizes the long term average throughput under the AWGN capacity utility function $\frac{1}{2}\log(1+x)$. The proposed policy is near optimal in the sense that it performs within constant multiplicative and additive gaps from the optimal solution that is independent of energy arrivals and battery sizes. This is extended to broadcast channels in \cite{baknina_online_bc}, multiple access channels in \cite{ozgur_online_mac, baknina_online_mac}, and systems with processing costs in \cite{baknina_online_su_proc, baknina_online_twc_proc} (for examples of earlier online approaches see, e.g., \cite{mitran-eh-online, dong-eh-online, elif-eh-online}).

In this paper, we generalize the approaches in \cite{ozgur_online_su} to work for general concave monotonically increasing utility functions for single-user channels. That is, we consider the design of online power control policies that maximize the long term average general utilities. One motivation for this setting is energy harvesting receivers. Since power consumed in decoding is modelled as a convex increasing function of the incoming rate \cite{kayaRxEH, yatesRxEH1,arafaJSACdec}, the rate achieved at the receiver is then a concave increasing function of the decoding power. In our setting, energy is harvested in packets that follow an i.i.d. distribution with amounts known causally at the transmitter. The transmitter has a finite battery to store its harvested energy. We first study the special case of Bernoulli energy arrivals that fully recharge the battery when harvested, and characterize the optimal online solution. Then, for the general i.i.d. arrivals, we show that the policy introduced in \cite{ozgur_online_su} performs within a constant multiplicative gap from the optimal solution for any general concave increasing utility function, for all energy arrivals and battery sizes. We then provide sufficient conditions on the utility function to guarantee that such policy is within a constant additive gap from the optimal solution.


\section{System Model and Problem Formulation}

We consider a single-user channel where the transmitter relies on energy harvested from nature to send its messages to the receiver. Energy arrives (is harvested) in packets of amount $E_t$ at the beginning of time slot $t$. Energy packets follow an i.i.d. distribution with a given mean. Our setting is {\it online:} the amounts of energy are known causally in time, i.e., after being harvested. Only the mean of the energy arrivals is known a priori. Energy is saved in a battery of finite size $B$.

Let $u$ be a differentiable, concave, and monotonically increasing function representing some general utility (reward) function, with $u(0)=0$ and $u(x)>0$ for $x>0$, and let $g_t$ denote the transmission power used in time slot $t$. By allocating power $g_t$ in time slot $t$, the transmitter  achieves $u\left(g_t\right)$ instantaneous reward. Denoting $\mathcal{E}^t\triangleq\{E_1,E_2,\dots,E_t\}$, a feasible online policy ${\bm g}$ is a sequence of mappings $\{g_t:~\mathcal{E}^t\rightarrow \mathbb{R}_+\}$ satisfying
\begin{align}
0\leq g_t\leq b_t\triangleq\min\{b_{t-1}-g_{t-1}+E_t,B\},\quad \forall t
\end{align}
with $b_1\triangleq B$ without loss of generality (using similar arguments as in \cite[Appendix B]{ozgur_online_su}). We denote the above feasible set by $\mathcal{F}$. Given a feasible policy ${\bm g}$, we define the $n$-horizon average reward as
\begin{align}
\mathcal{U}_n({\bm g})\triangleq\frac{1}{n}\mathbb{E}\left[\sum_{t=1}^nu\left(g_t\right)\right]
\end{align}
Our goal is to design online power scheduling policies that maximize the long term average reward subject to (online) energy causality constraints. That is, to characterize
\begin{align} \label{opt_main}
\rho^*\triangleq\max_{{\bm g}\in\mathcal{F}}\lim_{n\rightarrow\infty}\mathcal{U}_n({\bm g})
\end{align}

\section{Main Results}

In this section, we present the main results of this paper. We note that problem (\ref{opt_main}) can be solved by dynamic programming techniques since the underlying system evolves as a Markov decision process. However, the optimal solution using dynamic programming is usually computationally demanding with few structural insights. Therefore, in the sequel, we aim at finding relatively simple online power control policies that are provably within a constant additive and multiplicative gap from the optimal solution for all energy arrivals and battery sizes.

We assume that $E_t\leq B~\forall t$ a.s., since any excess energy above the battery capacity cannot be saved or used. Let $\mu=\mathbb{E}[E_t]$, where $\mathbb{E}[\cdot]$ is the expectation operator, and define
\begin{align} \label{eq_fraction_q}
q\triangleq\frac{\mathbb{E}[E_t]}{B}
\end{align}
Then, we have $0\leq q\leq1$ since $E_t\leq B$ a.s. We define the power control policy as follows \cite{ozgur_online_su}
\begin{align} \label{eq_ffp}
\tilde{g}_t=qb_t
\end{align}
That is, in each time slot, the transmitter uses a fixed fraction of its available energy in the battery. Such policies were first introduced in \cite{ozgur_online_su}, and coined {\it fixed fraction policies} (FFP). Clearly such policies are always feasible since $q\leq1$. Let $\rho\left(\tilde{{\bm g}}\right)$ be the long term average utility under the FFP $\{\tilde{g}_t\}$. We now state the main results.

\begin{lemma} \label{thm_ub}
The optimal solution of problem (\ref{opt_main}) satisfies
\begin{align}
\rho^*\leq u(\mu)
\end{align}
\end{lemma}

\begin{theorem} \label{thm_mult}
The achieved long term average utility under the FFP in (\ref{eq_ffp}) satisfies
\begin{align}
\frac{1}{2}\leq\frac{\rho\left(\tilde{{\bm g}}\right)}{u\left(\mu\right)}\leq1
\end{align}
\end{theorem}

We note that the results in Lemma~\ref{thm_ub} and Theorem~\ref{thm_mult} indicate that the FFP in (\ref{eq_ffp}) achieves a long term average utility that is within a constant multiplicative gap from the optimal solution that is equal to $\frac{1}{2}$. This result is proved in \cite{ozgur_online_su} for $u(x)=\frac{1}{2}\log(1+x)$. Here, we are generalizing it to work for any concave increasing function $u$ with $u(0)=0$.

Next, we state the additive gap results. We first define
\begin{align}
h_{\theta}(x)\triangleq u(\theta x)-u(x)
\end{align}
for some $0\leq\theta\leq1$, and define the following two classes of utility functions.

\begin{definition}[Utility Classes] \label{def_class}
A utility function $u$ belongs to class $(A)$ if $h_{\theta}(x)$ does not converge to 0 as $x\rightarrow\infty$, and belongs to class $(B)$ if $\lim_{x\rightarrow\infty}h_{\theta}(x)=0$.
\end{definition}

Now let us define the following function for $0<\theta<1$
\begin{align}
h(\theta)\triangleq\inf_xh_\theta(x)
\end{align}
whenever the infimum exists. Note that the infimum exists for class $(B)$ utility functions since $h_\theta(x)<0$ for $x>0$ by monotonicity of $u$, and $h_\theta(0)=0$. We state some properties of the function $h$ in the next lemma. The proof follows by monotonicity and concavity of $u$ and is omitted for brevity.
\begin{lemma} \label{thm_h}
$h(\theta)$ is non-positive, concave, and non-decreasing in $\theta$.
\end{lemma}

The next two theorems summarize the additive gap results for utility functions in classes $(A)$ and $(B)$ in Definition~\ref{def_class}.

\begin{theorem} \label{thm_add}
If $h(\theta)$ exists, and if
\begin{align}
r\triangleq(1-q)\lim_{t\rightarrow\infty}\frac{1-\lim_{x\rightarrow\bar{x}_{t+1}}u\left((1-q)^{t+1}x\right)/u(x)}{1-\lim_{x\rightarrow\bar{x}_t}u\left((1-q)^tx\right)/u(x)}<1
\end{align}
where $\bar{x}_t\in\arg \inf_xh_{(1-q)^t}(x)$; then the achieved long term average utility under the FFP in (\ref{eq_ffp}) satisfies
\begin{align}
u\left(\mu\right)+\alpha\leq\rho\left(\tilde{{\bm g}}\right)\leq u\left(\mu\right)
\end{align}
where $\alpha\triangleq\sum_{t=0}^\infty q(1-q)^th\left((1-q)^t\right)$ is finite.
\end{theorem}

\begin{theorem} \label{thm_add_asym}
For class $(B)$ utility functions, the achieved long term average utility under the FFP in (\ref{eq_ffp}) satisfies
\begin{align}
\lim_{\mu\rightarrow\infty}\rho\left(\tilde{{\bm g}}\right)=\rho^*
\end{align}
\end{theorem}

We note that the results in Lemma~\ref{thm_ub} and Theorem~\ref{thm_add} indicate that the FFP in (\ref{eq_ffp}) achieves a long term average utility, under some sufficient conditions, that is within a constant additive gap from the optimal solution that is equal to $\left|\sum_{t=0}^\infty q(1-q)^th\left((1-q)^t\right)\right|$. One can further make this gap independent of $q$ by minimizing it over $0\leq q\leq1$. We discuss examples of the above results in Section~\ref{sec_ex}, where we also comment on FFP performance under utility functions that do not satisfy the sufficient conditions in Theorem~\ref{thm_add}.

\section{Bernoulli Energy Arrivals}

In this section, we characterize the optimal solution of a special case of the energy arrival i.i.d. process: the Bernoulli process. Let $\{\hat{E}_t\}$ be a Bernoulli energy arrival process with mean $\mu$ as follows
\begin{align} \label{eq_bern_energy}
\hat{E}_t\in\{0,B\},~\text{with}~\mathbb{P}[\hat{E}_t=B]=p,~\text{and}~pB=\mu
\end{align}
where $\mathbb{P}[A]$ denotes the probability of $A$. Note that under such specific energy arrival setting, whenever an energy packet arrives, it completely fills the battery, and resets the system. This constitutes a {\it renewal}. Then, by \cite[Theorem 3.6.1]{ross_stochastic} (see also \cite{ozgur_online_su}), the following holds for any power control policy ${\bm g}$
\begin{align} \label{eq_renewal}
\lim_{n\rightarrow\infty}\hat{\mathcal{U}}_n({\bm g})&=\lim_{n\rightarrow\infty}\frac{1}{n}\mathbb{E}\left[\sum_{t=1}^nu\left(g_t\right)\right] \nonumber \\
&=\frac{1}{\mathbb{E}[L]}\mathbb{E}\left[\sum_{t=1}^Lu\left(g_t\right)\right]\quad\text{a.s.}
\end{align} 
where $\hat{\mathcal{U}}_n({\bm g})$ is the $n$-horizon average utility under Bernoulli arrivals, and $L$ is a random variable denoting the inter-arrival time between energy arrivals, which is geometric with parameter $p$, and $\mathbb{E}[L]=1/p$.

Using the FFP defined in (\ref{eq_ffp}) in (\ref{eq_renewal}) gives a lower bound on the long term average utility. Note that by (\ref{eq_bern_energy}), the fraction $q$ in (\ref{eq_fraction_q}) is now equal to $p$. Also, the battery state decays exponentially in between energy arrivals, and the FFP is
\begin{align} \label{eq_ffp_bern}
\tilde{g}_t&=p(1-p)^{t-1}B=(1-p)^{t-1}\mu
\end{align}
for all time slots $t$, where the second equality follows since $pB=\mu$. Using (\ref{eq_renewal}), problem (\ref{opt_main}) in this case reduces to
\begin{align}
\max_{{\bm g}}\quad&\sum_{t=1}^\infty p(1-p)^{t-1}u\left(g_t\right) \nonumber \\
\mbox{s.t.}\quad&\sum_{t=1}^\infty g_t\leq B,\quad g_t\geq0,\quad\forall t
\end{align}
which is a convex optimization problem. The Lagrangian is,
\begin{align}
\mathcal{L}=-\sum_{t=1}^\infty p(1-p)^{t-1}u\left(g_t\right)+\lambda\left(\sum_{t=1}^\infty g_t-B\right)-\sum_{t=1}^\infty\eta_tg_t
\end{align}
where $\lambda$ and $\{\eta_t\}$ are Lagrange multipliers. Taking derivative with respect to $g_t$ and equating to 0 we get
\begin{align}
u^\prime\left(g_t\right)=\frac{\lambda-\eta_t}{p(1-p)^{t-1}}
\end{align}
Since $u$ is concave, then $u^\prime$ is monotonically decreasing and $f\triangleq(u^\prime)^{-1}$ exists, and is also monotonically decreasing. By complementary slackness, we have $\eta_t=0$ for $g_t>0$, and the optimal power in this case is given by
\begin{align} \label{eq_bern_opt_pwr}
g_t=f\left(\frac{\lambda}{p(1-p)^{t-1}}\right)
\end{align}
and it now remains to find the optimal $\lambda$. We note by monotonicity of $f$, $\{g_t\}$ is non-increasing, and it holds that
\begin{align} \label{eq_bern_pos_pwr}
g_t=f\left(\frac{\lambda}{p(1-p)^{t-1}}\right)>0 \Leftrightarrow \lambda<p(1-p)^{t-1}u^\prime(0)
\end{align}
Hence, if $u^\prime(0)$ is infinite, then (\ref{eq_bern_pos_pwr}) is satisfied $\forall t$, and the optimal power allocation sequence is an infinite sequence. In this case, we solve the following equation for the optimal $\lambda$
\begin{align} \label{eq_bern_lambda_inf}
\sum_{t=1}^\infty f\left(\frac{\lambda}{p(1-p)^{t-1}}\right)=B
\end{align}
which has a unique solution by monotonicity of $f$.

On the other hand, for finite $u^\prime(0)$, there exists a time slot $N$, after which the second inequality in (\ref{eq_bern_pos_pwr}) is violated since $\lambda$ is a constant and $p(1-p)^{t-1}$ is decreasing. In this case the optimal power allocation sequence is only positive for a finite number of time slots $1\leq t\leq N$. We note that $N$ is the smallest integer such that
\begin{align} \label{eq_bern_fin_pwr}
\lambda\geq p(1-p)^Nu^\prime(0)
\end{align}
Thus, to find the optimal $N$ (and $\lambda$), we first assume $N$ is equal to some integer $\{2,3,4,\dots\}$, and solve the following equation for $\lambda$
\begin{align}
\sum_{t=1}^N f\left(\frac{\lambda}{p(1-p)^{t-1}}\right)=B
\end{align}
We then check if (\ref{eq_bern_fin_pwr}) is satisfied for that choice of $N$ and $\lambda$. If it is, we stop. If not, we increase the value of $N$ and repeat. This way, we reach a KKT point, which is sufficient for optimality by convexity of the problem \cite{boyd}. We note that for $u(x)=\frac{1}{2}\log(1+x)$ whose $u^\prime(0)$ is finite, \cite{ozgur_online_su} called $N$, $\tilde{N}$. We generalize their analysis for any concave increasing function $u$. This concludes the discussion of the optimal solution in the case of Bernoulli energy arrivals.

\section{General i.i.d. Energy Arrivals:\\Proofs of Main Results}

\subsection{Proof of Lemma~\ref{thm_ub}}

In this section, we derive the upper bound in Lemma~\ref{thm_ub} that works for all i.i.d. energy arrivals. Following \cite{ozgur_online_su} and \cite{baknina_online_su_proc}, we first remove the battery capacity constraint setting $B=\infty$. This way, the feasible set $\mathcal{F}$ becomes
\begin{align} \label{eq_feas_off}
\sum_{t=1}^ng_t\leq\sum_{t=1}^nE_t,\quad\forall n
\end{align}
Then, we remove the expectation and consider the offline setting of problem (\ref{opt_main}), i.e., when energy arrivals are known a priori. Since the energy arrivals are i.i.d., the strong law of large numbers indicates that $\lim_{n\rightarrow\infty}\frac{1}{n}\sum_{t=1}^nE_t=\mu$ a.s., i.e., for every $\delta>0$, there exists $n$ large enough such that $\frac{1}{n}\sum_{t=1}^nE_t\leq\mu+\delta$ a.s., which implies by (\ref{eq_feas_off}) that the feasible set, for such $(\delta,n)$ pair, is given by
\begin{align}
\frac{1}{n}\sum_{t=1}^ng_t\leq\mu+\delta\quad\text{a.s.}
\end{align}

Now fix such $(\delta,n)$ pair. The objective function is given by
\begin{align}
\frac{1}{n}\sum_{t=1}^nu(g_t)
\end{align}
Since $u$ is concave, the optimal power allocation minimizing the objective function is $g_t=\mu+\delta$, $1\leq t\leq n$ \cite{boyd} (see also \cite{jingP2P}). Whence, the optimal offline solution is given by $u(\mu+\delta)$. We then have $\rho^*\leq u(\mu+\delta)$. Since this is true $\forall\delta>0$, we can take $\delta$ down to 0 by taking $n$ infinitely large.

\subsection{Proof of Theorem~\ref{thm_mult}}

We first derive a lower bound on the long term average utility for Bernoulli energy arrivals under the FFP as follows
\begin{align} 
\lim_{n\rightarrow\infty}\hat{\mathcal{U}}_n(\tilde{\bm g})&\stackrel{(a)}{=}p\sum_{i=1}^\infty p(1-p)^{i-1}\sum_{t=1}^iu\left(\tilde{g}_t\right) \nonumber \\
&=\sum_{t=1}^\infty p(1-p)^{t-1}u\left((1-p)^{t-1}\mu\right) \label{eq_second_equality} \\
&\stackrel{(b)}{\geq}\sum_{t=1}^\infty p(1-p)^{2(t-1)}u\left(\mu\right) \nonumber \\
&=\frac{1}{2-p}u(\mu) \geq\frac{1}{2}u(\mu) \label{eq_mult_lb}
\end{align}
where $(a)$ follows by (\ref{eq_renewal}), $(b)$ follows by concavity of $u$ \cite{boyd}, and the last inequality follows since $0\leq p\leq1$. Next, we use the above result for Bernoulli arrivals to bound the long term average utility for general i.i.d. arrivals under the FFP in the following lemma; the proof follows by concavity and monotonicity of $u$, along the same lines of \cite[Section VII-C]{ozgur_online_su}, and is omitted for brevity.
\begin{lemma} \label{thm_bern_iid}
Let $\{\hat{E}_t\}$ be a Bernoulli energy arrival process as in (\ref{eq_bern_energy}) with parameter $q$ as in (\ref{eq_fraction_q}) and mean $qB=\mu$. Then, the long term average utility under the FFP for any general i.i.d. energy arrivals, $\rho(\tilde{\bm g})$, satisfies
\begin{align}
\rho(\tilde{\bm g})\geq\lim_{n\rightarrow\infty}\hat{\mathcal{U}}_n(\tilde{\bm g})
\end{align}
\end{lemma}

Using Lemma~\ref{thm_ub}, (\ref{eq_mult_lb}), and Lemma~\ref{thm_bern_iid}, we have
\begin{align}
\frac{1}{2}u(\mu)\leq \rho(\tilde{\bm g})\leq\rho^*\leq u(\mu)
\end{align}

\subsection{Proof of Theorem~\ref{thm_add}}

By Lemma~\ref{thm_ub} and Lemma~\ref{thm_bern_iid}, it is sufficient to study the lower bound in the case of Bernoulli arrivals. By (\ref{eq_second_equality}) we have
\begin{align}
\lim_{n\rightarrow\infty}\hat{\mathcal{U}}_n(\tilde{\bm g})&=\sum_{t=1}^\infty p(1-p)^{t-1}u\left((1-p)^{t-1}\mu\right) \nonumber \\
&\stackrel{(c)}{\geq}\sum_{t=1}^\infty p(1-p)^{t-1}\left(u\left(\mu\right)+h\left((1-p)^{t-1}\right)\right) \nonumber \\
&=u(\mu)+\sum_{t=0}^\infty p(1-p)^th\left((1-p)^t\right) \nonumber \\
&\triangleq u(\mu)+\alpha
\end{align}
where $(c)$ follows since $h(\theta)$ exists, and is by definition no larger than $h_\theta(x)$, $\forall x,\theta$. Now to check whether $\alpha$ is finite, we apply the ratio test to check the convergence of the series $\sum_{t=0}^\infty (1-p)^th\left((1-p)^t\right)$. That is, we compute
\begin{align}
r&\triangleq\lim_{t\rightarrow\infty}\left|\frac{(1-p)^{t+1}h\left((1-p)^{t+1}\right)}{(1-p)^th\left((1-p)^t\right)}\right| \nonumber \\
&=(1-p)\lim_{t\rightarrow\infty}\frac{\inf_x1-u\left((1-p)^{t+1}x\right)/u(x)}{\inf_x1-u\left((1-p)^tx\right)/u(x)}
\end{align}
where the second equality follows by definition of $h$. Next, we replace $\inf_x$ by $\lim_{x\rightarrow\bar{x}_t}$ since $\bar{x}_t\in\arg\inf h_{(1-p)^t}(x)$, and take the limit inside (after the 1). Finally, if $r<1$ then $\alpha$ is finite; if $r>1$ then $\alpha=-\infty$; and if $r=1$ then the test is inconclusive and one has to compute $\lim_{T\rightarrow\infty}\sum_{t=0}^T p(1-p)^th\left((1-p)^t\right)$ to get the value of $\alpha$.

\subsection{Proof of Theorem~\ref{thm_add_asym}}

For utility functions of class $(B)$, we have $\lim_{x\rightarrow\infty}u(\theta x)-u(x)=0$. Thus, $\forall\epsilon>0$ there exists $\bar{\mu}$ large enough such that
\begin{align}
u\left((1-p)^{t-1}\mu\right)>u\left(\mu\right)-\epsilon,\quad\forall\mu\geq\bar{\mu}
\end{align}
whence, for Bernoulli energy arrivals we have
\begin{align}
\lim_{n\rightarrow\infty}\hat{\mathcal{U}}_n(\tilde{\bm g})&=\sum_{t=1}^\infty p(1-p)^{t-1}u\left((1-p)^{t-1}\mu\right) \nonumber \\
&\geq u\left(\mu\right)-\epsilon,\quad\forall\mu\geq\bar{\mu}
\end{align}
It then follows by Lemma~\ref{thm_ub} and Lemma~\ref{thm_bern_iid} that
\begin{align}
\rho^*\geq\rho\left(\tilde{{\bm g}}\right)\geq u\left(\mu\right)-\epsilon\geq\rho^*-\epsilon,\quad\forall\mu\geq\bar{\mu}
\end{align}
and we can take $\epsilon$ down to 0 by taking $\mu$ infinitely large.

\section{Examples and Discussion} \label{sec_ex}

In this section we present some examples to illustrate the results of this work. We first show that the utility function $u(x)=\frac{1}{2}\log(1+x)$ considered in \cite{ozgur_online_su} belongs to class $(A)$. Indeed we have $h_\theta^\prime(x)=\frac{\theta-1}{2(1+\theta x)(1+x)}$, which is negative for all $0<\theta<1$, and therefore $h_\theta(x)$ is decreasing in $x$ and does not converge to 0. We then show that the sufficient conditions of Theorem~\ref{thm_add} are satisfied: $h(\theta)$ exists, and is equal to $\lim_{x\rightarrow\infty}\frac{1}{2}\log\frac{1+\theta x}{1+x}=\frac{1}{2}\log(\theta)$; $r=1-q$ and hence the gap $\alpha$ is finite. Furthermore, \cite{ozgur_online_su} showed that minimizing $\alpha$ over all $q$ gives a constant additive gap, independent of $q$, that is equal to $0.72$.

Next, we note that all bounded utility functions belong to class $(B)$. These are functions $u$ where there exists some constant $M<\infty$ such that $u(x)\leq M,~\forall x$. Examples for these include: $u(x)=1-e^{-\beta x}$ for some $\beta>0$, and $u(x)=x/(1+x)$. To see that these functions belong to class $(B)$, observe that $\lim_{x\rightarrow\infty}u(x)=M$ by monotonicity of $u$, and hence $\lim_{x\rightarrow\infty}u(\theta x)-u(x)=0$. We also note that class $(B)$ is not only inclusive of bounded utility functions. For example, the unbounded function $u(x)=\sqrt{\log(1+x)}$ satisfies $\lim_{x\rightarrow\infty}\sqrt{\log(1+\theta x)}-\sqrt{\log(1+x)}=0$ and therefore belongs to class $(B)$. For such unbounded functions in class $(B)$, the FFP is not only within a constant additive gap of the optimal solution, it is asymptotically optimal as well, as indicated by Theorem~\ref{thm_add_asym}.

Note that one can find a (strict) lower bound on $h(\theta)$ for some utility functions if it allows more plausible computation of $\alpha$, or if $h(\theta)$ itself is not direct to compute. For instance, for any bounded utility function $u$, the following holds: $h(\theta)\geq(\theta-1)M$, where $M$ is the upper bound on $u$. To see this, observe that by concavity of $u$ and the fact that $u(0)=0$ we have $\inf_xu(\theta x)-u(x)\geq(\theta-1)\sup_xu(x)$. This gives $\alpha\geq\sum_{t=0}^\infty q(1-q)^t\left((1-q)^t-1\right)M$, which is no smaller than $-\frac{1}{2}M$ if we further minimize over $q$. Another example is $u(x)=\frac{1}{2}\log\left(1+\sqrt{x}\right)$, which belongs to class $(A)$. We observe that $h(\theta)$ in this case is lower bounded by $\frac{1}{2}\log(\theta)$. Hence, this function admits an additive gap no larger than $0.72$ calculated in \cite{ozgur_online_su} for $u(x)=\frac{1}{2}\log(1+x)$.

Finally, we note that the conditions of Theorem~\ref{thm_add} are only sufficient for the FFP defined in (\ref{eq_ffp}) to be within an additive gap from optimal. For instance, consider $u(x)=\sqrt{x}$. This function belongs to class $(A)$ as $h_\theta(x)$ does not converge to 0. In fact, $h_\theta(x)$ is unbounded below and $h(\theta)$ does not exist. This means that any FFP of the form $\tilde{g}_t=\theta b_t$, for any choice of $0<\theta<1$, is {\it not} within a constant additive gap from the upper bound $\sqrt{\mu}$. However, there exists another FFP (with a different fraction than $q$ in (\ref{eq_fraction_q})) that is {\it optimal} in the case of Bernoulli arrivals. Since $u^\prime(0)=\infty$, we use (\ref{eq_bern_lambda_inf}) to find the optimal $\lambda$, where $f(x)=1/(4x^2)$, and substitute in (\ref{eq_bern_opt_pwr}) to get that the optimal transmission scheme is {\it fractional:} $g_t =\hat{p}\left(1-\hat{p}\right)^{(t-1)}B,~\forall t$, where the transmitted fraction $\hat{p}\triangleq1-(1-p)^2$. This shows that one can pursue near optimality results under an FFP by further optimizing the fraction of power used in each time slot, and comparing the performance directly to the optimal solution instead of an upper bound. While in this work, we compared the lower bound achieved by the FFP to a universal upper bound that works for all i.i.d. energy arrivals.

\bibliographystyle{unsrt}

\end{document}